\documentclass[12pt,a4paper]{article}
\usepackage[latin1]{inputenc}
\usepackage{amsmath}
\usepackage{amsfonts}
\usepackage{amssymb}
\usepackage{graphicx}
\usepackage{calc}
\usepackage{wrapfig}
\usepackage[a4paper,margin=1.0in]{geometry}
\usepackage{subfig}
\usepackage{cite}
\usepackage{hyperref}
\newcommand{\ee}{\ensuremath{\mathrm{e}^+\mathrm{e}^-\,}}
\newcommand{\ie}{{\it{i.e.}}}
\newcommand{\eg}{{\it{e.g.}}}

\begin{document}
\title{Status of the undulator-based ILC positron source}
\author{F. Dietrich$^1$, 
G. Moortgat-Pick$^2$,
S. Riemann$^1$\thanks{corresponding author, sabine.riemann@desy.de}, P. Sievers$^3$,  A. Ushakov$^2$\\
\\ $~$ 
 \normalsize $^1$\textit{Deutsches Elektronen-Synchrotron (DESY), Platanenallee 6, D-15738 Zeuthen}\\
 \normalsize $^2$\textit{University of Hamburg, Luruper Chaussee 149, D-22761 Hamburg} \\
 \normalsize $^3$\textit{CERN, CH-1211 Geneva 23, Switzerland}}
\maketitle
\begin{abstract}
\noindent
The  design of the positron source for the International Linear Collider (ILC) is still under consideration. The baseline design plans to  use the electron beam for the positron poduction before it goes to the IP. The high-energy electrons pass a long helical undulator and  generate an intense circularly polarized photon beam which hits a thin conversion target to produce \ee{} pairs.  The resulting positron beam is longitudinally polarized which provides an important benefit for  precision physics analyses.  
In this paper the status of the design  studies is presented with focus on ILC250. In particular, the target design and  cooling as well as issues of the optical matching device are important for the  positron yield. Some possibilities to optimize the system are discussed.
\end{abstract}.
\section{Introduction}\label{sec:intro}

The positron production for a high-energy linear \ee collider is a challenge;  about $1.3\times 10^{14}$ positrons per second  are required at the ILC collision point for nominal luminosity. The efficiency of positron production in a conversion target together with the capture acceleration of the positrons is not high. The load on the target and other source components as well as the radiation aspects are demanding issues for the optimization of positron sources for projects like ILC or CLIC. 

The ILC positron production~\cite{ref:TDR} is based on a  long helical undulator passed by the high energy electron beam to create  an intense circularly polarized  photon beam. The photon beam hits a thin conversion target to produce electron-positron pairs; the resulting positron beam is longitudinally polarized. The target material is currently specified as Ti6Al4V. The target is designed as wheel of 1\,m diameter  spinning with 2000 rounds per minute in vacuum. This rotation speed is necessary to distribute the heat load from the intense, narrow photon beam to a larger volume  of the target material during one ILC pulse.   The capture system behind the target consists of an optical matching device followed by  accelerator structures in a solenoidal  0.5\,T field. 
Here, the parameters for the undulator-based positron source are considered in more detail with focus on ILC250.  One important issue is  the target cooling. Since water cooling turned out to be extremely complicated, cooling by thermal radiation is  currently the favored option. Further, the choice of the capture optics is decisive for the positron yield.
The status for the optimization of the positron source parameters is presented. Since the polarized positron beam is an outstanding feature of the ILC undulator source,  in section~\ref{sec:e+pol} this benefit for physics measurements is shortly summarized. The following sections consider the design issues in detail. Focus is the load on the  conversion target  and its cooling.  Since the load on the target depends on the efficiency of the  capture system, also the positron yield is considered depending on the optical macthing device. 
Some possibilities  to optimize the source parameters are discussed.

It should be mentioned that alternatively  an electron beam of few GeV can be used to create \ee pairs in a thick target. In reference \cite{ref:edriven} the application of such system for the ILC is suggested; the status and progress are described in \cite{ref:e+WG}. However, the resulting positron beam will be unpolarized.  
\section{Polarization of positrons}\label{sec:e+pol}
 
Future high-energy \ee{} linear colliders will probe the Standard Model and physics beyond with excellent precision.  
Electroweak interactions do not conserve parity, so beam polarization is essential to measure and to disentangle phenomena.
 High degrees of electron beam  polarization are possible; the ILC e$^-$ beam will be at least 80\% polarized.
Since the generation of an intense (polarized) positron beam is a challenge, simultaneously polarized e$^-$ and e$^+$  beams are under discussion since many years.
Without going in details as  physics processes and their analyses, the benefit of polarized positron beams is given by the following reasons (see also references~\cite{MoortgatPick:2005cw,Karl:2017xra,Robert-Thesis}:
\begin{itemize}
\item
There are  4 combinations of e$^+$ and e$^-$ helicity states in the collision of high-energy electrons and positrons. 
Only with both beams polarized  each of these  initial state combinations can be explicitly realized in a collider. 
\item
With the 'right' helicity combination of initial states a higher effective luminosity  is achieved: ${\cal L}_{\rm eff}/{\cal L}=1-(1-P_{e^-}P_{e^+})$.
A  higher  number of specific events is achieved in shorter running time. For example,
assuming $P_\mathrm{e^-}=90\%$ and  $P_\mathrm{e^+}=30\%$ the effective luminosity can be almost a factor  1.3 higher than without positron polarization. 
The availability of both beams polarized reduces therefore the required running time by one third.
\item
The suppression of background is crucial for precision measurements. With polarized beams the desired initial states can be enhanced or suppressed. This improves the discrimination and control of background processes.  
\item
Polarized beams provide a high flexibility to evaluate systematic effects.
It is very difficult to detect and correct time-dependent effects, correlations or a bias in the polarimeter measurement.  If both beams are polarized, such systematic effects can be much better controled, and their impact on the uncertainty of observables  can be substantially reduced down to negligable values.
\item
In case of deviations from the Standard Model predictions, polarization of both beams
enhances significantly the possibility to confirm the existence of a new phenomenon: High precision,  flexible configuration of initial states and a larger number of independent observables could even allow to unravel underlying physics.
\item
An independent determination of beam polarization and left-right asymmetries is possible but only  if 
both beams are polarized. 
\end{itemize}
One should keep in mind that also the zero polarization of an unpolarized positron beam must be confirmed  
to avoid any bias in the physics analyses~\cite{Fujii:2018mli}.
\\
All these arguments suggest that positron polarization is crucial already for ILC250. Detailed analyses demonstrate that with 2\,ab$^{-1}$ and polarized e$^+$ and e$^-$ beams a great Higgs physics program is offered: Many couplings can be measured with an uncertainty of 1\% or better~\cite{Fujii:2018mli,updateTDR}.  This probes new physics phenomena complementary to the LHC.  To achieve this precision with polarized  electrons only, 5\,ab$^{-1}$ are required.

\section{The positron source parameters for ILC250}\label{sec:params}
\begin{figure}[htbp]
\center
 \includegraphics*[width=120mm]{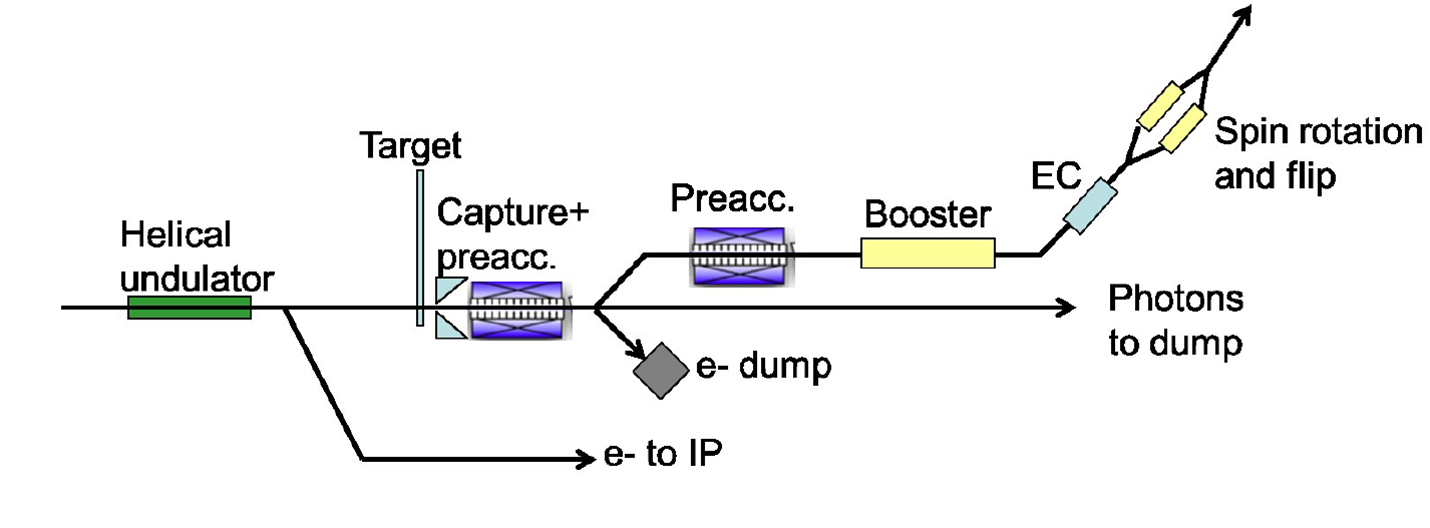}
  \caption{Sketch of the undulator-based ILC positron source.} 
 \label{fig:source}
\end{figure}

The ILC positron source is located at the end of the main linac. It consists of the helical undulator with maximum active length of 231\,m, the conversion target made of Ti6Al4V, the optical matching device and the capture optics, acceleration, energy and bunch compression, spin rotation and spin flipper as shown in figure~\ref{fig:source} and reference~\cite{ref:TDR}. 
Since the photon energy and yield, and hence the positron yield depend strongly on the electron energy, the source performance  has to be studied and optimized for each centre-of-mass energy. Goal is a positron yield of 1.5 e$^+$/e$^-$ at the damping ring.

\subsection{Superconducting helical undulator}\label{sec:helund}
Central part of the positron source is the superconducting helical undulator. A prototype was manufactured and tested in UK~\cite{ref:undproto}. It consists of a 4\,m long cryomodule with two 1.75\,m long undulator modules. The undulator period is $\lambda_{\mathrm{u}}=11.5\,$mm and the maximum B field is 0.86\,T corresponding to a  value $K=0.92$. As described in the TDR, with 132 undulator modules (66 cryomodules) an active undulator length of 231\,m is reached. With the  quadrupoles foreseen every 3 cryomodules the undulator system reaches a total length of 320\,m. 
The given undulator period, $\lambda_{\mathrm{u}}$, and the maximum field on axis, $B_0$, define the possible parameter range  for the positron source, \ie{} the undulator $K$ value, $K\propto \lambda_{\mathrm{u}} B_0$.
The efficiency of positron generation in the target depends on the pair production cross section and hence on the photon energy. The cut-off for the first harmonic is related to the electron energy $E_\mathrm{e}$, $K$ and $\lambda_{\mathrm{u}}$ by
\begin{equation}
E_{1 \gamma} \propto  \frac{E_{\mathrm{e}}}{\lambda_{\mathrm{u}}(1+K^2)}\,,\label{eq:Egamma}
\end{equation}
i.e. lower K values increase the photon energy.
The number of photons created per undulator length is 
\begin{equation}
N_{\gamma} \propto  \frac{K^2}{\lambda_{\mathrm{u}}(1+K^2)}\,,\label{eq:Ngamma}
\end{equation}
implying that low $K$ values result in less photons.   
If the electron beam of the ILC250 machine is used for the positron production, a high $K$ value and the full active length of the undulator are required. 
\\
Also the beam spot size on the target depends on the opening angle of the photon beam,
\begin{equation}
\theta_{\gamma} \propto  \frac{\sqrt{1+K^2}}{\gamma}\,,\label{eq:theta}
\end{equation}  
and it is very small even at a large distance from the undulator.  
The narrow photon beam causes a high peak energy deposition density (PEDD) at the target. To prevent overheating during one ILC pulse, the target is spinning with 100\,m/s circumferential speed.
 Both,  PEDD as well as the average power deposited in the target vary for different $E_\mathrm{cm}$ and depend on the  distance  between target and undulator. 
In previous studies the interplay of parameters has been studied for different centre-of-mass energies~\cite{ref:LCWS17proc,ref:e+WG,ref:posipol16-AU,ref:AU-250GeVthickness,ref:AU-OMD}.
Most of these studies assumed a pulsed flux concentrator (FC) as optical matching device (OMD). A promising prototype study for the FC was performed by LLNL~\cite{ref:Gronberg-FC}. However, detailed studies showed that load at the inner part of the flux concentrator front side is high, at least too high for ILC250~\cite{ref:AU-OMD}. This is mainly caused by the larger opening angle of the photon beam which is $\propto 1/\gamma$ 
and the wider distribution of the shower particles downstream the target due to the lower photon energy at ILC250. 
To resolve this problem, the drift space between the middle of undulator and the target was reduced to 401\,m. In addition, masks can be included to protect the OMD. Alternatively, a quarter wave transformer should be used which has a larger aperture than the FC; further details can be found in reference~\cite{ref:AU-OMD}.
Table~\ref{tab:sourcepar} presents an overview of the relevant parameters for the studies in this paper. The load on the target was simulated for FC and QWT. In both cases the positron beam  polarization is 30\%.
However, the positron yield depends strongly on the magnetic field assumed for the simulations. 
The numbers in table~\ref{tab:sourcepar} given for the QWT suppose an optimised shape of the B field; the maximum field of 1.04\,T is achieved at distance of ~8mm after the target exit instead of about 3.5\,cm in the design  given in reference~\cite{ref:GaiLiu-OMD}. More details are given in section~\ref{sec:OMD}.
\begin{table}[h]
\begin{center}
\renewcommand{\baselinestretch}{1.2}
\begin{tabular}{|lc|cc|}
\hline
                        &     &   FC & QWT\\ \hline
electron beam energy    & GeV &  \multicolumn{2}{c|}{126.5}\\
undulator active length &  m  & \multicolumn{2}{c|}{231} \\ 
space from middle of undulator to target & m & \multicolumn{2}{c|}{401} \\ \hline
undulator K             &     & 0.85& 0.92 \\
photon yield per m undulator & $\gamma$/(e$^- \, $m) & 1.70 &  1.95 \\
photon yield            & $\gamma$/e$^-$     &392.7&  450.4\\ 
photon energy (1$^\mathrm{st}$ harmonic) & MeV & 7.7  & 7.2\\ 
average photon energy                  & MeV & 7.5  & 7.6 \\
average photon beam power              &  kW & 62.6 & 72.2\\
average power deposited in target       & kW &  1.94 & 2.2 \\
rms photon beam spot size on target ($\sigma$) & mm &  1.2 &1.45 \\
PEDD in target per pulse (100\,m/s)       & J/g & 61.0 & 59.8\\
\hline 
\end{tabular}
\caption{\label{tab:sourcepar}Summary of the source performance parameters for ILC250 with  1312 bunches per pulse. The pulse repetition rate is 5Hz. The numbers are shown for a decelerating capture field.  See also references~\cite{ref:TDR,ref:posipol16-AU,ref:AU-250GeVthickness,ref:AU-OMD}. }
\end{center}
\end{table}

\section{Positron target wheel}\label{sec:wheel}

\subsection{Energy deposition}\label{sec:Edep}
\begin{figure}[htbp]
\begin{tabular}{lr}
 \includegraphics*[width=75mm]{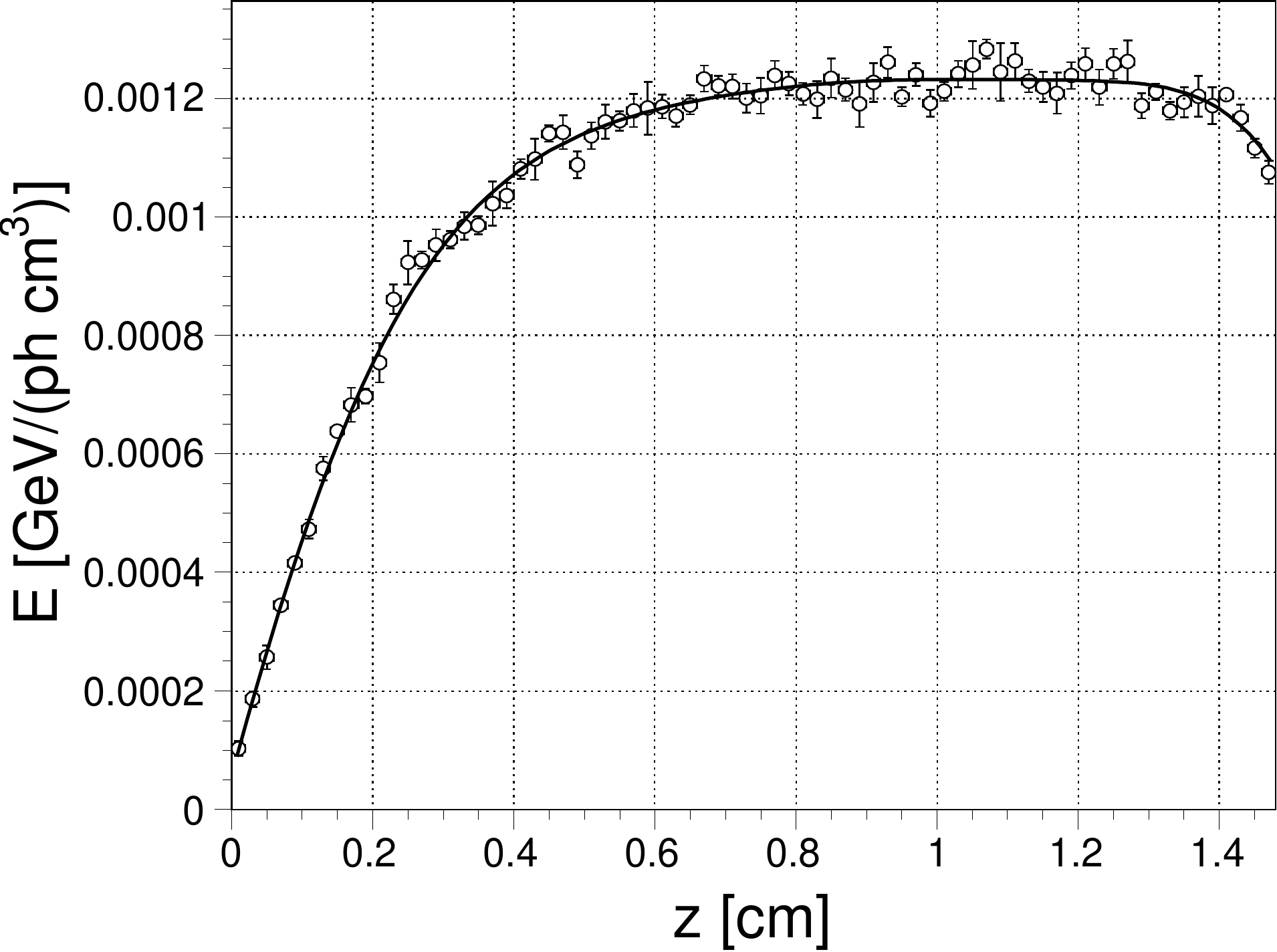}&
 \includegraphics*[width=75mm]{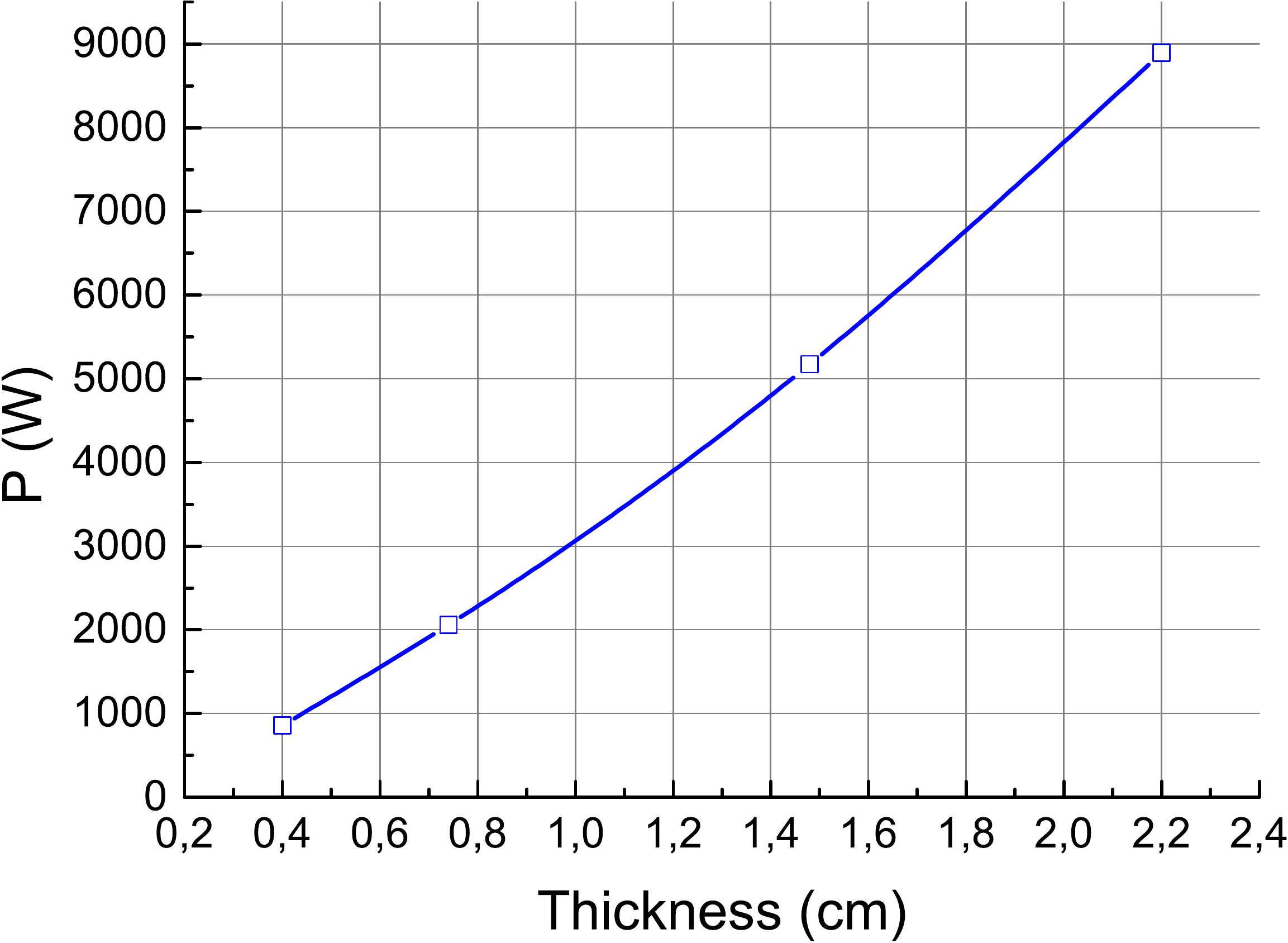}
\end{tabular}
  \caption{Energy deposition in the target for ILC250. {\it{Left:}} Energy deposition per photon of a 231\,m long undulator in the target along z. {\it{Right:}} Total energy deposition in the target depending on the thickness~\cite{ref:AU-250GeVthickness}. }
 \label{fig:Edep}
\end{figure}

The photon beam hits the spinning target of Ti6Al4V, the photons undergo the pair production process in the field of nuclei, electrons and positrons are generated and exit the target. The energy deposition in the target as well as the number and  energy of the exiting particles depend on the target thickness. For ILC500 a target thickness of 0.4 radiation length is recommended as described in the TDR. 
The energy deposition by a 120\,GeV electron beam in the target was simulated with FLUKA~\cite{ref:FLUKA}; the results are shown in figure~\ref{fig:Edep}~\cite{ref:AU-250GeVthickness}. The energy deposition along z and also the positron yield are almost constant between 7\,mm and 16\,mm target thickness but the power deposited in the target increases. 
Taking into account the cooling of the spinning target, a target thickness of 7\,mm is optimum.

\subsection{Cooling by thermal radiation}\label{sec:radcool}
As discussed in previous studies~\cite{ref:TDR} the average energy deposition in the ILC positron target is about $2-7\,$kW depending on the drive beam energy in the undulator, the target thickness and the luminosity (nominal or high). 
For ILC250, the average energy deposition in the target is 2\,kW. 
\\
Since the initial investigations of the wheel, involving leak tight rotating vacuum seals and water cooling showed major problems~\cite{ref:Gronberg-FC,ref:Gronberg-posipol13}, an alternative technical solution was brought up to ensure the heat radiation as well as the safe rotation of 2000\,rpm by magnetic bearings~\cite{ref:sievers-posipol14,ref:sievers-posipol16}. This proposal was intensively investigated and many studies followed which are resumed here in this paper.\\
Few kW can be extracted by radiation cooling if the radiating surface is large enough and the heat distributes fast enough from the area of incident  beam to a large radiating surface. 
Following the Stefan-Boltzmann law, 
\begin{equation}
P = \sigma_0 \varepsilon_\mathrm{eff} A (T^4-T_0^4)\,,\label{eq:T4}
\end{equation}
with $\sigma_0 =5.67\times 10^{-12}$\,W/(cm$^2\times$ K$^4$), a radiating surface $A=0.36\,$m$^2$ is required  to remove 2\,kW if the average temperature is $T_\mathrm{ave}=400^\circ$C and  $\varepsilon_\mathrm{eff}=0.5$.
 For comparison: the area of a  target rim with outer diameter  $r_o = 51\,$cm and inner diameter $r_i=45\,$cm is 0.36\,m$^2$ taking into account front and back side.
With other words: The wheel spinning in vacuum can radiate the heat to a stationary  cooler opposite to the wheel surface. It is easy to keep the stationary cooler at room temperature by water cooling.
But it is crucial for the design that the  heat distributes from the volume heated by the photon beam  to a larger surface area. 
The thermal conductivity of Ti6Al4V is low, $\lambda = 0.068\,$W/(cm\,K) at room temperatures and  0.126\,W/(cm\,K) at 540$^\circ$C. The heat capacity is $c=0.58\,$J/(g\,K) at room temperature and 0.126J/(g\,K) at 540$^\circ$C~\cite{ref:T-dep-parameters}. Although the wheel rotation frequency can be adjusted so that each part of the target rim is hit after 6-8 seconds, this time is not sufficient to distribute the heat load almost uniformly over a large area. For example: Following $s=\sqrt{\lambda t/\rho c}$, the heat propagates  $s\approx 4$\,mm during 6 seconds in Ti6Al4V.  The heat is accumulated in the rim and the  highest temperatures are located in a relatively small region around the beam path. 

\subsection{Wheel design options}\label{sec:w-design}
Two options for the target wheel  are currently  under consideration:
\begin{enumerate}
\item
The target wheel is a  disk with the required target thickness. 
\item
The target wheel consists of a  rim made of of the target material  which is connected to a radiator with large surface made of material with good heat conductivity. 
\end{enumerate}
In case (1) the radiating surface is limited, and due to the low thermal conductivity the gradient along the radius is large. The thermal radiation is quite low from the inner surface with low radii.  So, option (1) is  recommended for lower energy deposition. 
\\
In case (2) the radiating surface can be easily increased by fins. The target rim needs a minimum size to cover the electromagnetic shower for the pair-production; it will have a substantially higher temperature than the radiator. The fins will be worked from material with high thermal conductivity to speed-up the heat extraction. Such construction could be necessary in case of higher energy deposition, \eg{}  in case of high luminosity or polarization upgrade. 
\\
To reduce high thermal stress along the rim in the beam path area as expected for high average temperatures, the target could be manufactured in sectors which can expand. In a disc this can be realized by radial expansion slots. 
For the final construction of the target wheel FEM design  studies are necessary  to ensure a long-term operation followed by systematic tests using a mock-up. 
Here, for ILC250 option (1) is considered.  
 For the FEM studies all material parameters were taken into account temperature dependent. They are summarized in  references~\cite{ref:LCWS17proc,ref:e+WG} and taken from~\cite{ref:Ti-par,ref:ATI,ref:matweb,ref:T-dep-parameters} 
 
\subsection{Driving mechanism and bearing}\label{sec:mech}
The wheel design is determined by the energy deposition and cooling efficiency. For the final construction also the bearing, the drive motor etc. are important. This is not considered here since first  the load and the cooling specifications must be worked out before the engineering design will be finalized.  Magnetic bearings, used for fly wheels for energy storage, for vacuum pumps and for Fermi Choppers have been developed, and are available on the market (SKF, Kernforschungszentrum Juelich). Usually, they base on permanent magnet technology. They can be adapted to the operating conditions of the rotating Titanium wheel for positron production. 
Breidenbach {\it{et al.}}~\cite{ref:breidenbach}  have studied a bearing, based on electro-magnetic coils.
Both solutions should be feasible for the target wheel but further R\&D is necessary.

 The heating of the target yields a non-uniform temperature distribution and stress within the wheel. At a first glance~\cite{ref:FS-imbalances}, the corresponding deformation due to expansion does not yield imbalances of the spinning wheel.  
So far, the dynamic effects have not yet considered in detail. Comprehensive simulations are planned to study them in order to prepare a reliable wheel design.

  
\subsection{Safety issues}\label{sec:safety}
The energy stored in the wheel is 
\begin{equation}
E_\mathrm{wheel} = 0.5 J \omega^2\,,
\end{equation}
where $J$ is the moment of inertia. Assuming a full disk of 52\,cm radius and 2000rpm, about 72\,kJ are stored in the  wheel considered here for ILC250. 
Appropriate housing is required. Due to the short distance between OMD and  target  a  protection of the OMD against mechanical crash of the target seems impossible.

\section{Load  distribution in the target}\label{sec:Load}
\subsection{Temperature distribution}\label{sec:Tdistr}
\begin{figure}[htbp]
\center
 \includegraphics*[width=150mm]{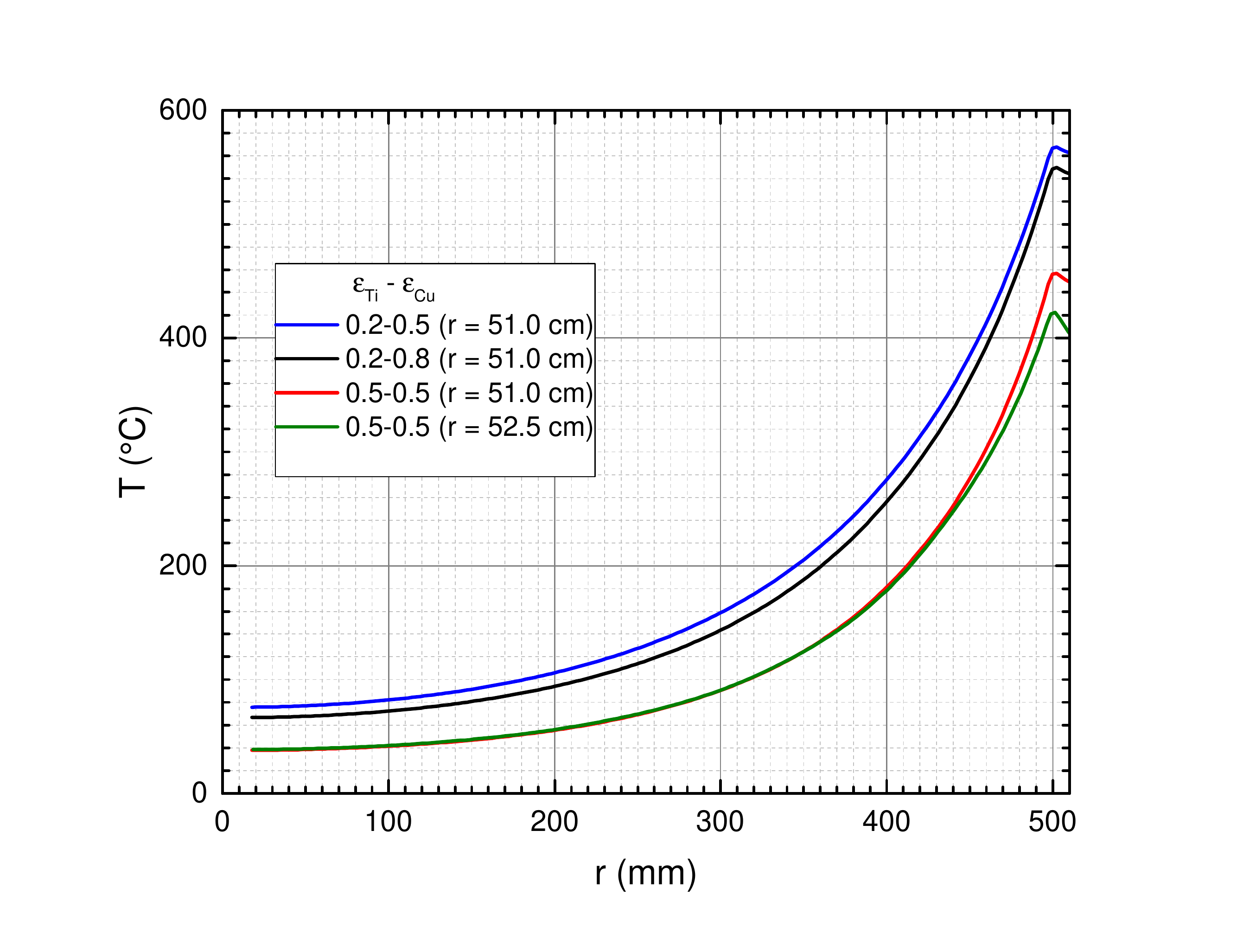}
  \caption{Radial average  temperature distribution in the target wheel without  radial expansion slots for various emissivities of target and cooler (copper). The beam hits the target at  a radius of 50\,cm. The outer wheel radius is 51\,cm and also 52.5\,cm.}
 \label{fig:resTall}
\end{figure}

The temperature  distribution in the target wheel determines the stress development in the target.  
At elevated temperatures the material will expand.  Since the  temperature gradient along the radius is large the highest thermomechanical stress is expected in the rim region 
where the beam impinges.
\\
Simulations with ANSYS~\cite{ref:ansys} were performed to study the temperature distribution and the corresponding thermal stress. The  temperature dependence of the material parameters (thermal conductivity, heat capacity) was taken into account (see also reference~\cite{ref:LCWS17proc}). 
The target is assumed as disk with thickness 7\,mm. 
The temperaure distribution was studied for various emissivities of target and cooler  surface. 
\\
The radial temperature distribution in the disc is shown in figure~\ref{fig:resTall}. It summarizes representative radial temperature profiles depending on the emissivities of the target and cooler material for a solid disc of 51\,cm and 52.5\,cm radius; the beam hits the target at a radius of 50\,cm. Larger wheel  radii increase slightly the radiative area in the hot rim region and decrease the maximum temperature. \\  
We also tested the influence of the distance between target surface and cooler which affects the radiating geometry in the hot rim region. It was found that this influence on the temperature  is almost negligible.\\ 
The emissivity of the Ti6Al4V  target sample used in the irradiation experiment at MAMI was measured to $\varepsilon \approx 0.5$. So it is expected that at least target and cooler surfaces
 with emissivities of 0.5 are feasable. This would result in an effective emissivity of  $\approx 0.33$ for the thermal radiation. As shown in    figure~\ref{fig:resTall}, in such case the maximum average temperature is roughly 460$^\circ$C. 
An optimization of the emissivities by surface processing or coating should be possible. Such scenarios are not yet taken into account and let room for improvement of the target cooling performance. 

\subsection{Cyclic load}\label{sec:cyclic}
In addition to the  high average temperatures in the rim region the photon beam creates cyclic load in the target which happens every 6-8 seconds at the same position, depending on the wheel revolution frequency. 
The photon beam  causes an instantaneous temperature rise which adds  up to roughly 80-100\,K (nominal luminosity) to the average temperature. Also for low emissivities the resulting peak temperature occurs only locally for short time and  does not exceed 600--650$^\circ$C. 

\begin{figure}[htbp]
\center
 \includegraphics*[width=100mm]{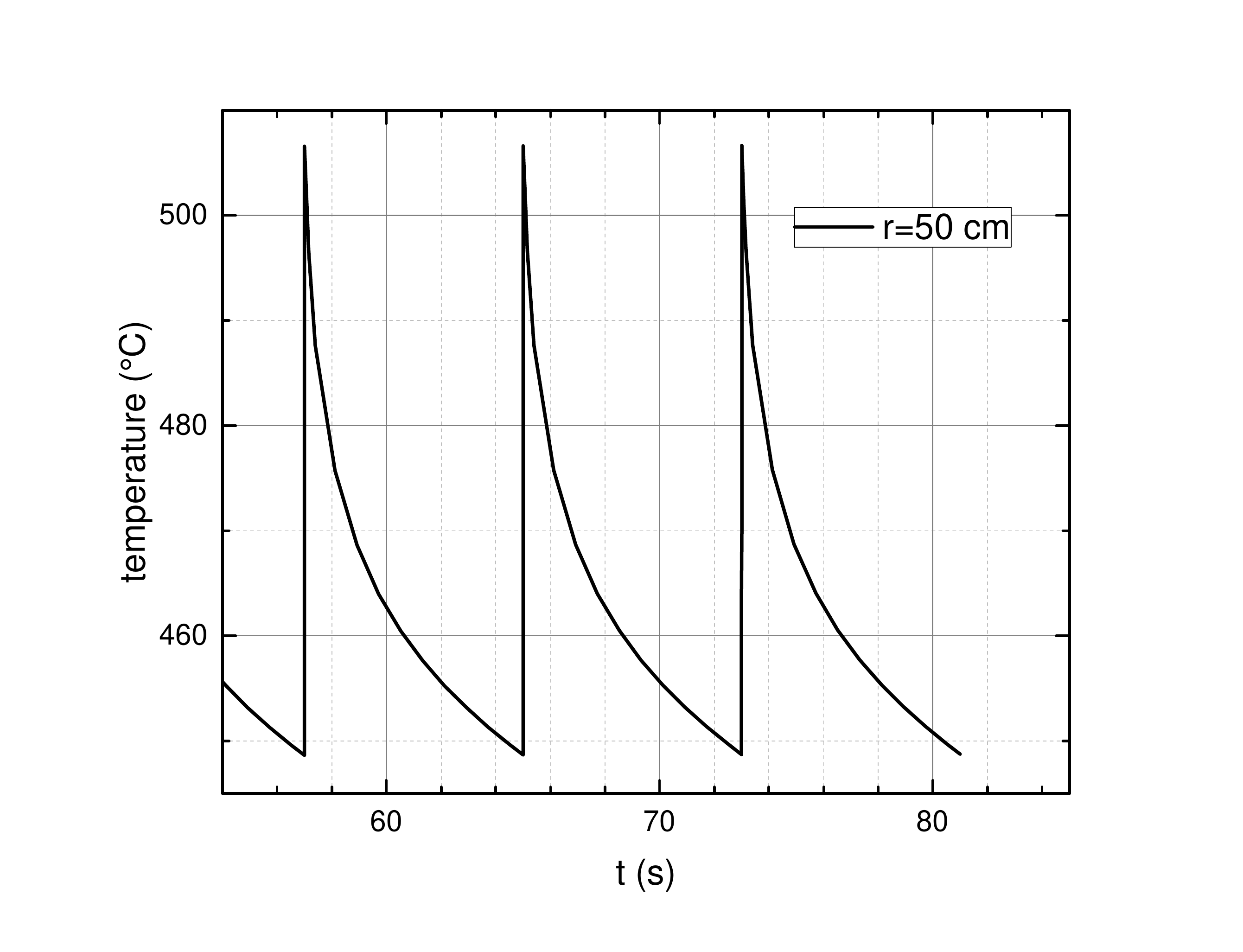}
  \caption{Temperature evolution in the target along the beam path with photon beam pulses. The effective emissivity is $\varepsilon_\mathrm{eff} = 0.33$.  }
 \label{fig:T-cycles}
\end{figure}

The cyclic temperature in the target rim caused by the beam impact is shown in figure~\ref{fig:T-cycles} for an effective emissivity of $\varepsilon_\mathrm{eff} = 0.33$ ($\varepsilon_\mathrm{Ti}=\varepsilon_\mathrm{Cu} = 0.5$). The cyclic load creates peak temperatures up to 510$^\circ$C; the maximum average temperature is 460$^\circ$C.

It was a very important result from experimental tests described in section~\ref{sec:mami} 
to see that no serious damage was obtained even up to temperatures close to the phase transition point (almost 1000$^\circ$C.).
However, the experimental test  did not take into account any load from a moving target.

\subsection{Stress in the target wheel}\label{sec:rotation}
\begin{figure}[htbp]
\center
 \includegraphics*[width=120mm]{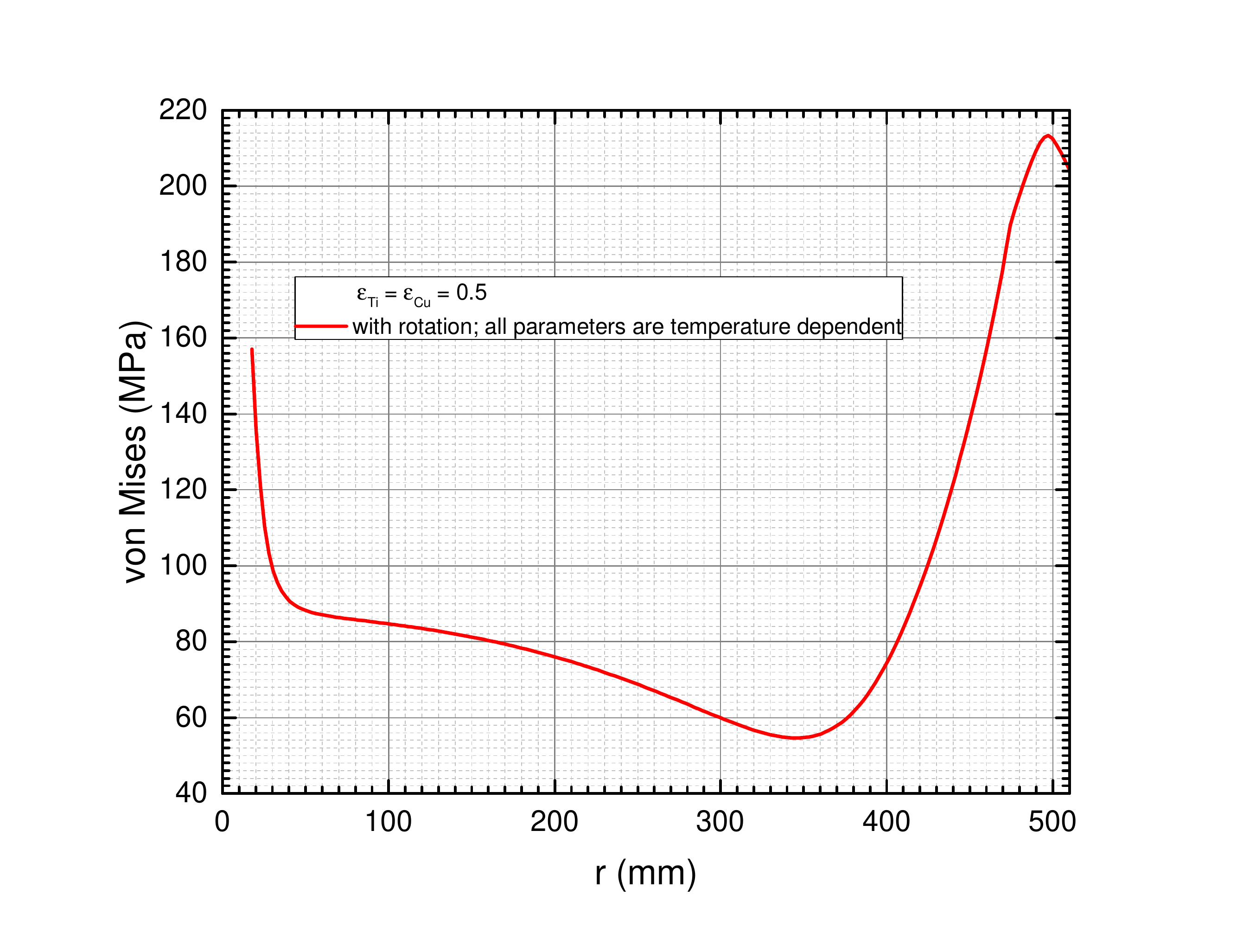}
  \caption{Radial distribution of average von Mises stress  in the full disc target wheel without radial expansion slots. The  wheel rotates with 2000\,rpm.} 
 \label{fig:wheel-stress}
\end{figure}

\begin{figure}[htbp]
\begin{flushleft}
\begin{tabular}{lr}
\hspace{-1cm} 
 \includegraphics*[width=90mm]{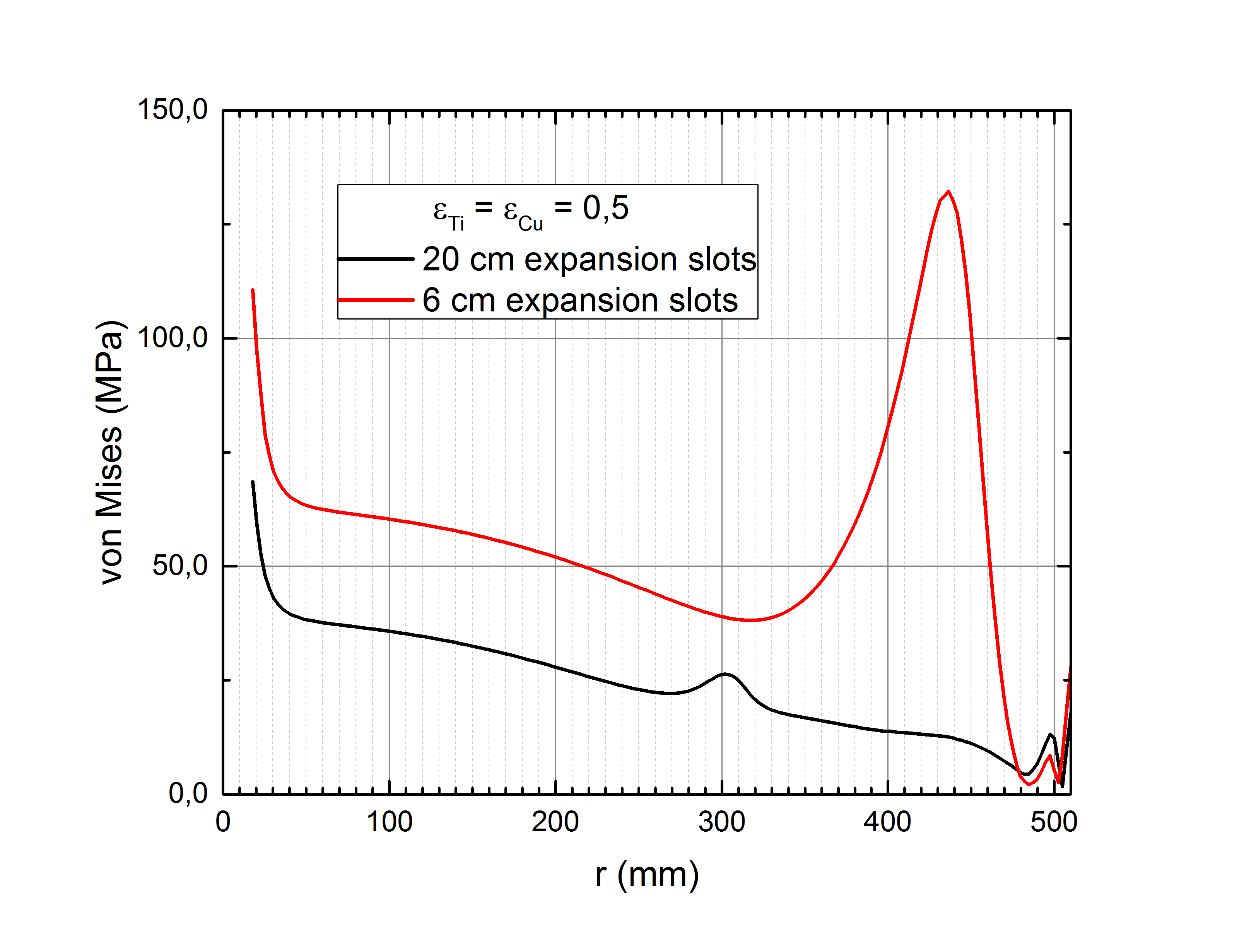}&
\hspace{-1cm} 
\includegraphics*[width=90mm]{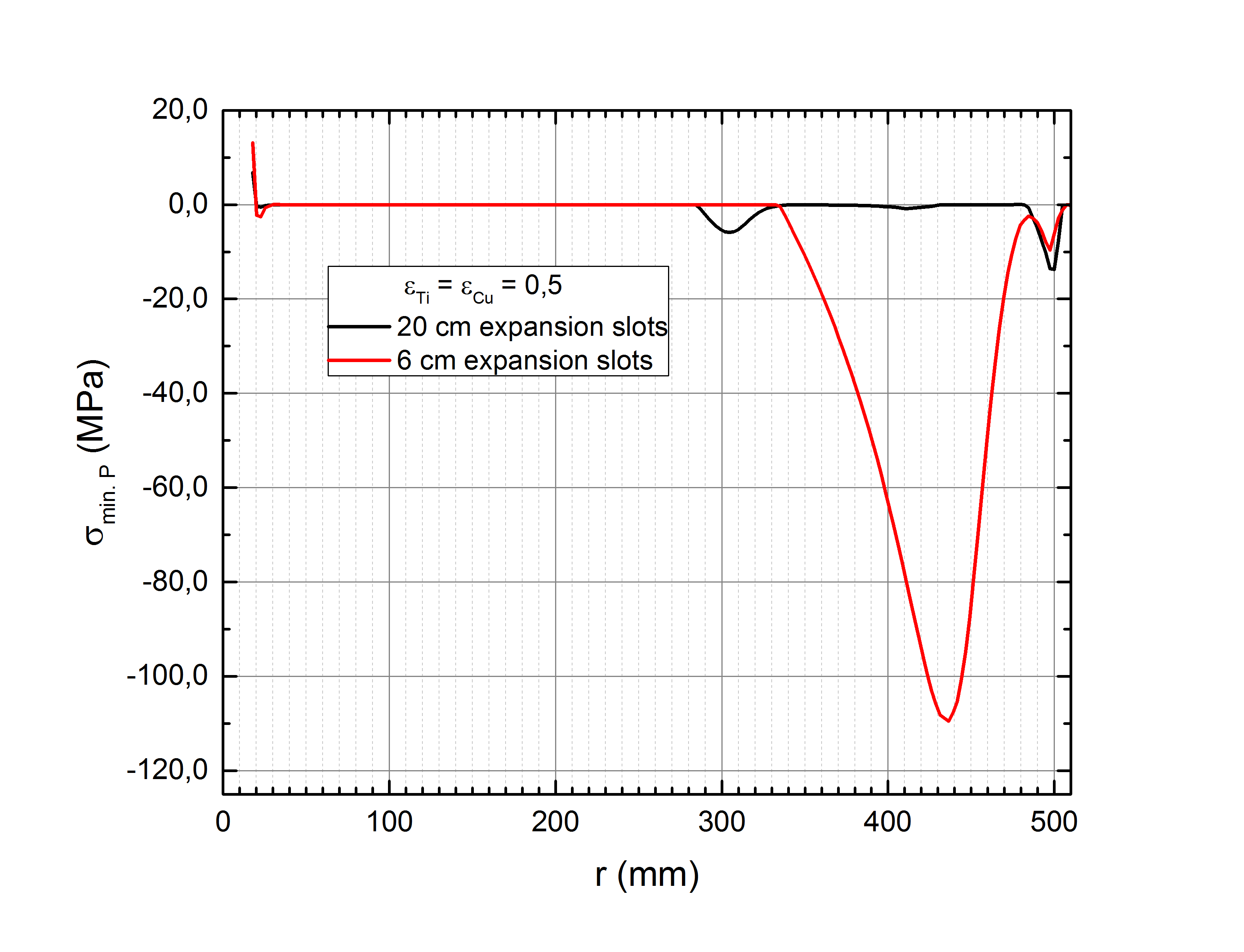}
\end{tabular}
  \caption{Radial distribution of average stress in the rotating target wheel with radial expansion slots of 6\,cm (red) and 20\,cm (black) length. {\it{Left:}} von Mises stress. {\it{Right:}} Minimal principal stress.} 
 \label{fig:wheel-stress-slots}
\end{flushleft}
\end{figure}

The high average temperature in the target rim as well as the cyclic load induce high stress due to thermal expansion.
To reduce the thermal stress in the target, expansion slots can be cut into the wheel. The temperature distribution remains almost unchanged but the stress is substantially reduced (see also~\cite{ref:LCWS17proc}).
For the studies of the stress development in the target wheel radial expansion slots of 6\,cm and 20\,cm were assumed; the distance between two slots corresponds to the beam path during one ILC pulse. 
If a wheel with such slots is operated the wheel rotation frequency has to be synchronized with the beam pulses. This  will  avoid that the photon beam hits a gap resulting in a short dip in the luminosity during one pulse. 
\\ 
The target rotation with $\omega\approx 200 $\,Hz increases the stress in the wheel in radial ($\sigma_\mathrm{r}$) and tangential ($\sigma_\mathrm{H}$) direction. Following classical treatment given in text books~\cite{ref:timoshenko,ref:dubbel} for forces in spinning wheels, 
the stress at the radius $r$ is given by
\begin{eqnarray}
\sigma_\mathrm{H} &=& \frac{3+\nu}{8}\rho \omega^2\left( 1-\frac{r^2}{r_o^2}\right)\left( 1-\frac{r_i^2}{r^2}\right) \\
\sigma_\mathrm{r} &=& \frac{3+\nu}{8}\rho \omega^2\left( 1+\frac{r_i^2}{r_o^2} +  \frac{r_i^2}{r^2}-\frac{1+3\nu}{3+\nu}\frac{r^2}{r_o^2}\right) 
\end{eqnarray}
with the inner and outer radius $r_i$ and $r_o$, the density $\rho$ and Poisson's ratio $\nu$. 
The target rotation increases the  stress in the hot target region  only little:
for a wheel with $r_o=52\,$cm the radial stress at the beam path radius ($r=50\,$cm) is 1.4\,MPa, the hoop stress along the beam path is 8.7\,MPa. 
The maximum hoop stress occurs at the inner wheel radius and is about 40\,MPa. The maximum radial stress is located at $r=\sqrt{r_o r_i}$, \ie{} at $r \approx~10 ...16\,$cm depending on $r_i=2...5\,$cm. In this region the temperature is about 100$^\circ$C. For the simple wheel geometry without expansion slots these results agree well with that of ANSYS simulations. 
A summary of the stress distributions simulated with ANSYS for a wheel without or  with expansion slots is given in figures~\ref{fig:wheel-stress} and~\ref{fig:wheel-stress-slots}. In case of expansion slots the highest stress is obtained at the radius where the slots end. The stress in figure~\ref{fig:wheel-stress-slots} is given along a radial line in the middle between two slots.
\\
The average thermal stress  and the cyclic load by the beam impact add up.  Maximum values are reached without expansion slots; they amount to about 300\,MPa. 
The stress in the wheel is compressive. 

\subsubsection{Expansion slots}\label{sec:slots}
\begin{figure}[htbp]
\center
\includegraphics*[width=95mm,height=50mm]{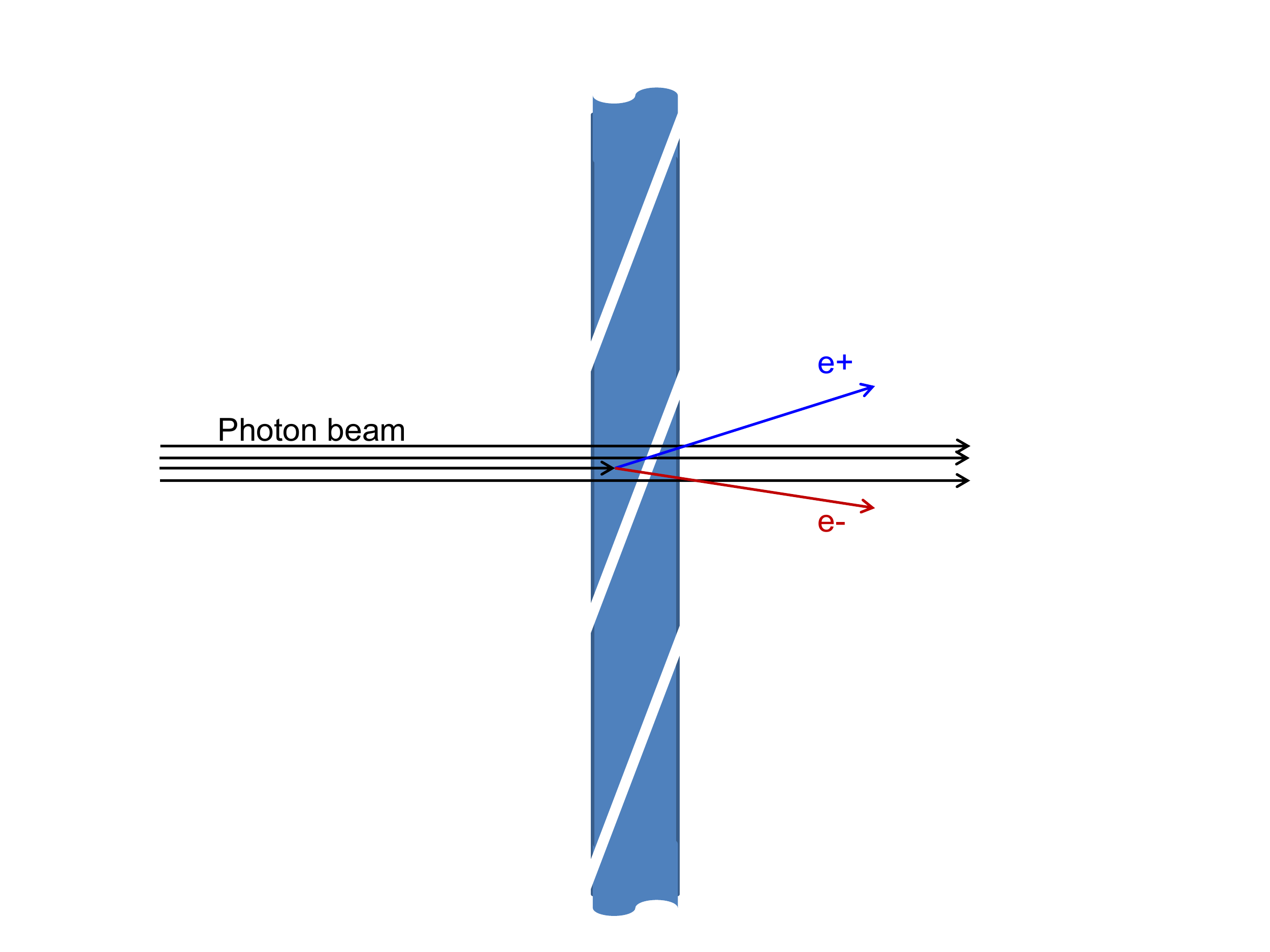}
  \caption{Top view on a piece of the target wheel to show a possible arrangment of expansion slots. The photon beam passes always the same target thickness if the wheel rotates. }
 \label{fig:slots}
\end{figure}

As mentioned, expansion slots reduce substantially the stress in the target rim. However, if the distance between the slots is not sunchronized with the bunch length and rotation speed of the target wheel, the positron production is not  constant during one bunch train. Many but narrow slots, \ie{} substantially smaller than the beam spot, would avoid the synchronisation of wheel rotation frequency and incident photon beam pulses. But fluctuations in the  luminosity will remain. 
Inclined slots as shown in figure~\ref{fig:slots} could be a solution: the photon beam passes always the constant target thickness; the fluctuations in the positron production should be negligable. 
Further studies are necessary to evaluate the  benefit and optimum shape of expansion slots. In particular, these studies have to take into account the mechanical stability at elevated temperatures. 

\subsection{Target wheel cooling and OMD}\label{sec:T+omd}
The optical matching  device should be installed as close as possible to the target. This could cause a magnetic field  in the target which induces  eddy currents in the spinning wheel.  These currents increase the target temperature~\cite{ref:sievers-eddy} and could slightly drag the wheel rotation; more details are discussed in section~\ref{sec:OMD}.
\\
 The OMD has a radius of about $25-30\,$cm and  occupies a  quite large part of the target surface. 
 When the hot target passes this OMD area, the heat radiates to the OMD and not to the cooler surfaces. The results presented above do not take into  account the somewhat lower cooling efficiency in the OMD.  Either the OMD can be cooled excellently or the target temperatures increase slightly. The heat load in the OMD due to  the radiating target has to be taken into account for its final design, also  with regard to the particle shower which  deposits energy in the OMD too.

\section{Target material test}\label{sec:mami}
It is an essential question whether the target material stands the high thermal and mechanical load. The instantaneous load  repeats up to $\approx 3\times 10^{6}$ times at the same target position within 1 year of ILC operation, so that also the fatigue limits must be considered. 
Further, the target is passed by a photon  beam which could change the material properties by damaging the material structure. To study this complex of questions, experimental tests were performed.   
 To simulate a load corresponding to that expected at the ILC positron target, the  electron beam (14\,MeV and 3.5\,MeV) of the Mainz Mictrotron (MAMI) injector was used.  
It was managed to focus the rms spot size of the 14\,MeV electron beam on target  to $\sigma \approx 0.2\,$mm for  the chopped 50\,$\mu$A cw electron beam with pulse length of 2\,ms~\cite{ref:mami-beam}. The 2\,ms pulse produced in Ti6Al4V samples (1\,mm and 2\,mm thickness) load conditions as expected at the ILC target. With  a repetition rate of 100\,Hz up to $6.6\times 10^6$ load cycles were generated correspnding to about 2 years ILC running.  All samples survived the irradiation procedure without damage visible by eyes. The grain structure was modified in samples that  reached maximum temperatures near to the phase transition value. For the details see reference~\cite{ref:MAMI-IPAC17}. It was concluded that the operation of a positron target consisting of Ti6Al4V is possible if the maximum temperature corresponds to the recommended operation temperature and exceeds only locally for short time this level up to  about $700^\circ$C. 
These results were confirmed irradiating thin samples of about $200-500$\,$\mu$m thickness with the 3.5\,MeV electron beam. 
The  strongly focused beam allowed a temperature rise by $160-350$\,K within a $1-5$\,ms pulse. Due to the fast repetition rate of the pulses (up to 100-140\,Hz) the average temparature  at  the considered location of the material reached $400-600^\circ$C.
The detailed analys including structural changes ist still omgoing. But with a laser microscope dimensional changes or visible modifications have not been obtained. Only if the the full power of the electron beam   was shortly directed for few seconds to the sample so that immediately temperatures near and above $1000^\circ$C were achieved buckling was abserved. Wholes in the smples have not been obtained. 
Although the studies are not yet finalized it is clear that high temperature Ti alloys as Ti6Al4V are  well suited as target material. 
Very high temperature Titanium alloys, for example Ti SF-61 could be an alternative as suggested in reference~\cite{ref:breidenbach}; the Titanium alloy SF-61 stands higher operation temperatures than Ti6Al4V.

\subsection{Load limits in Ti6Al4V}
High temperature titanium alloys are ductile materials. The  temperatures reached in the positron target or on potential exit windows made of Ti alloys rise the question for the limit to plastic deformations. Plastic deformations must be avoided since they could cause imbalances in the spinning wheel arrangement.  

\begin{figure}[htbp]
\begin{flushleft}
\begin{tabular}{lr}
\hspace{-1cm} 
 \includegraphics*[width=90mm]{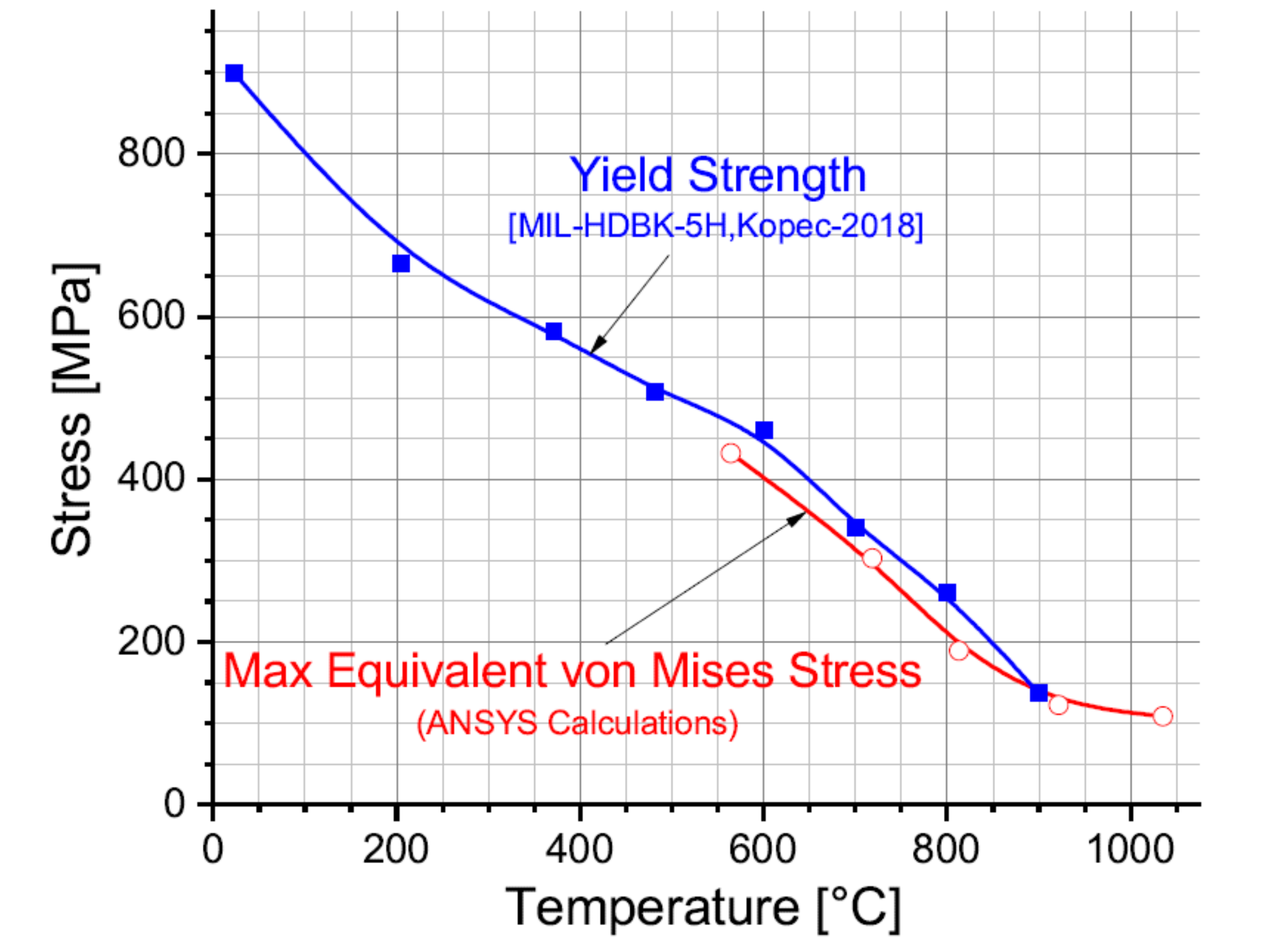}&
\hspace{-1cm} 
\includegraphics*[width=90mm]{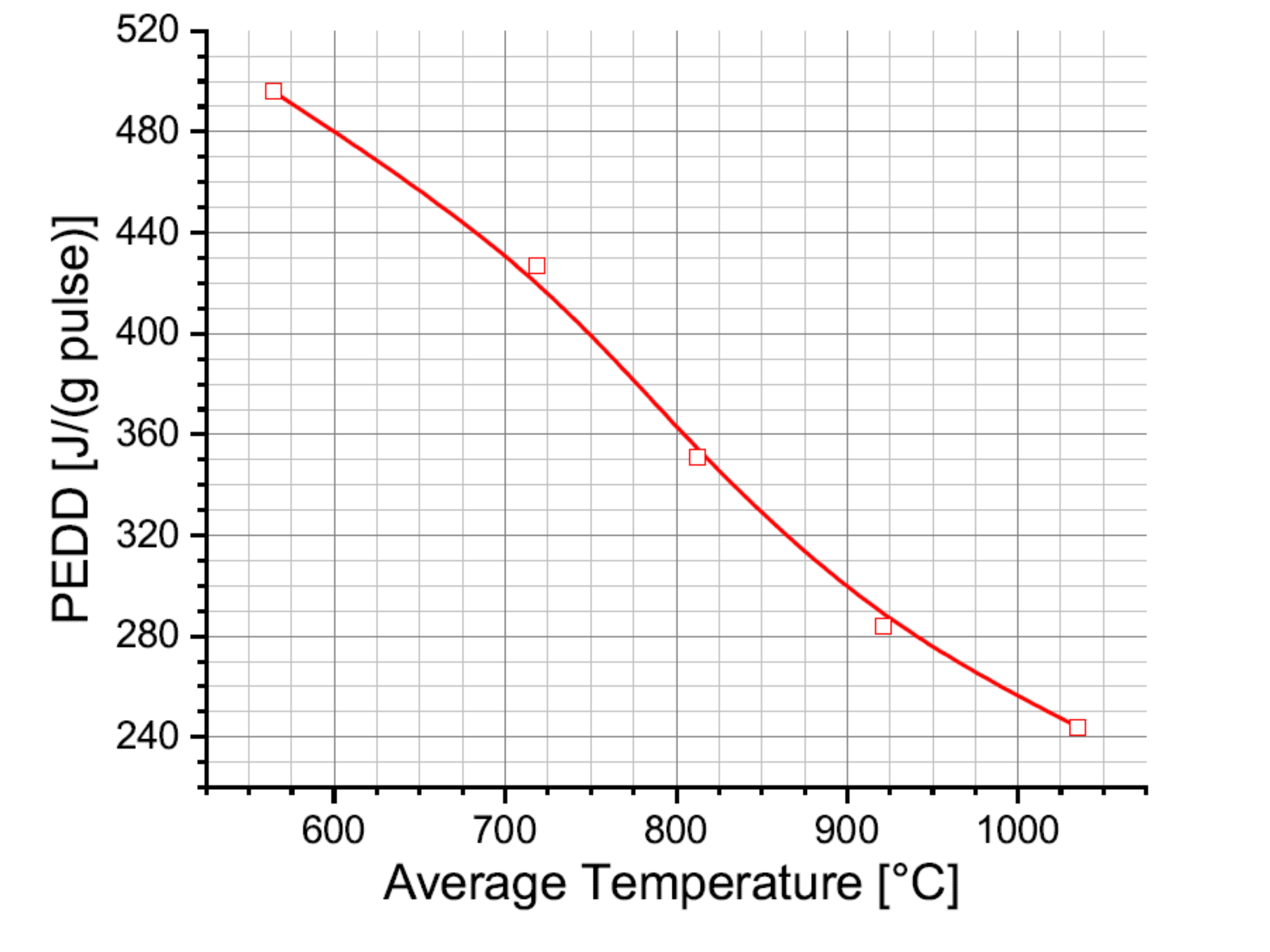}
\end{tabular}
  \caption{Maximum allowed thermal stress (left) and allowed peak energy deposition density (right), PEDD,  as function of the the temperature in Ti6Al4V~\cite{ref:AU-posipol18}.} 
 \label{fig:stress-limits}
\end{flushleft}
\end{figure}

To study this question, temperature dependent elasticity and hardening data were added to the
material properties in ANSYS and compared with the results of the material tests at MAMI in Mainz.  This allows to  estimate temperature and stress at which the material starts to deform plastically.
The simulations have shown that the maximum equivalent stress for
elastic deformation due to a particle beam in Ti6Al4V is approximately 10\% below the
yield strength.
At a temperature of $\approx600^\circ$C the transition from elastic to plastic deformation occurs
if the equivalent von Mises stress reaches 400\,MPa.
At higher temperatures the limits go down quickly, for example, at $800^\circ$C
average temperature, the  equivalent stress has to be  below 200\,MPa.
The results of this study are summarized in figure~\ref{fig:stress-limits}, more details can be found in reference~\cite{ref:AU-posipol18}.
Figure~\ref{fig:stress-limits} presents the maximum allowed thermal stress as well as the allowed peak energy deposition density (PEDD) as function of the temperature in Ti6Al4V. \\
The load expected in the target wheel (see section~\ref{sec:rotation}) is safely below the limit for plastic deformation. 
\\
Further studies are needed to include also creep effects. Also the impact of potential radiation damage must be understood.

\section{Positron yield and OMD}\label{sec:OMD}
\begin{figure}[h]
\centering
\begin{tabular}{lr}
 \includegraphics*[width=75mm]{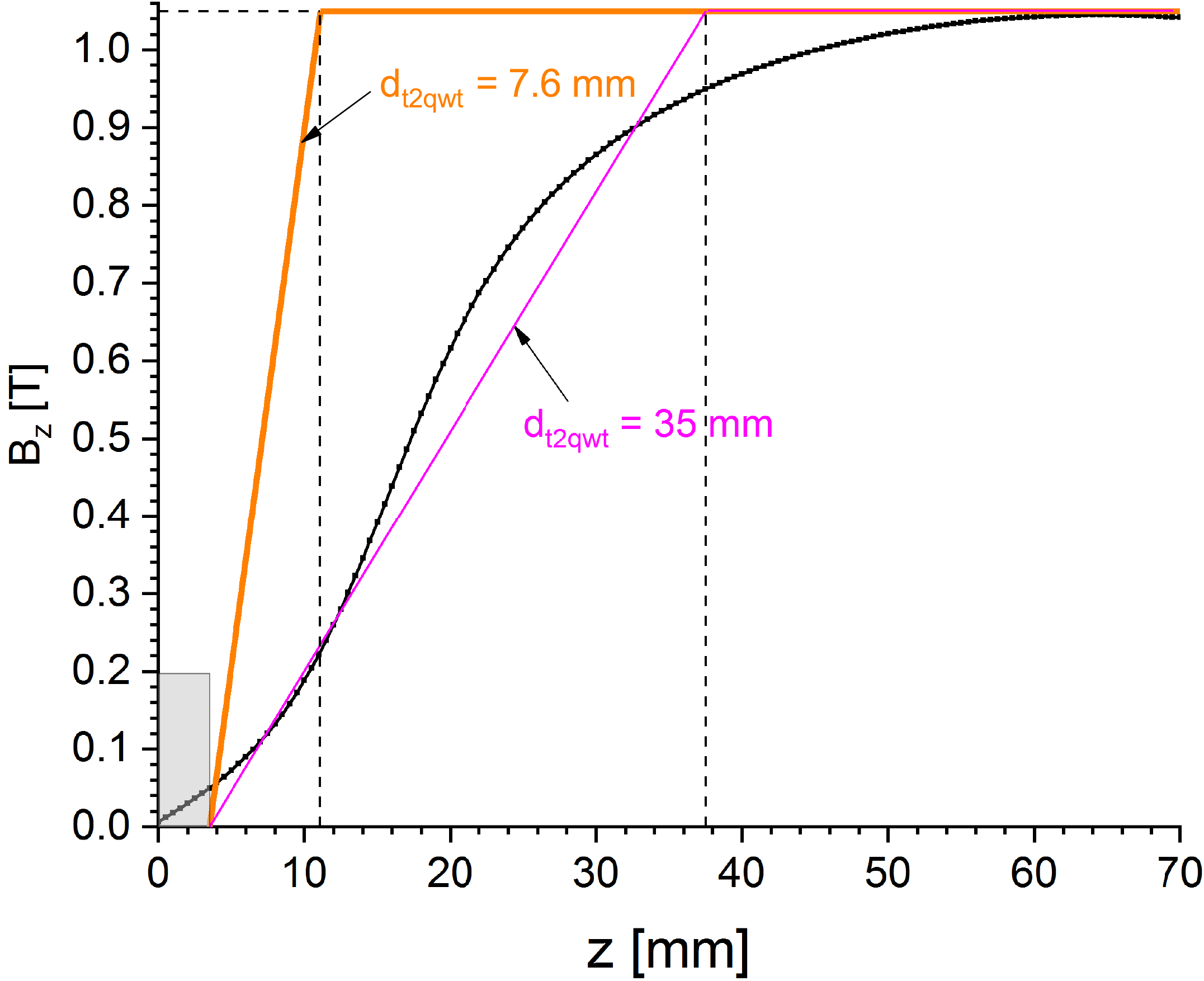}&
 \includegraphics*[width=75mm]{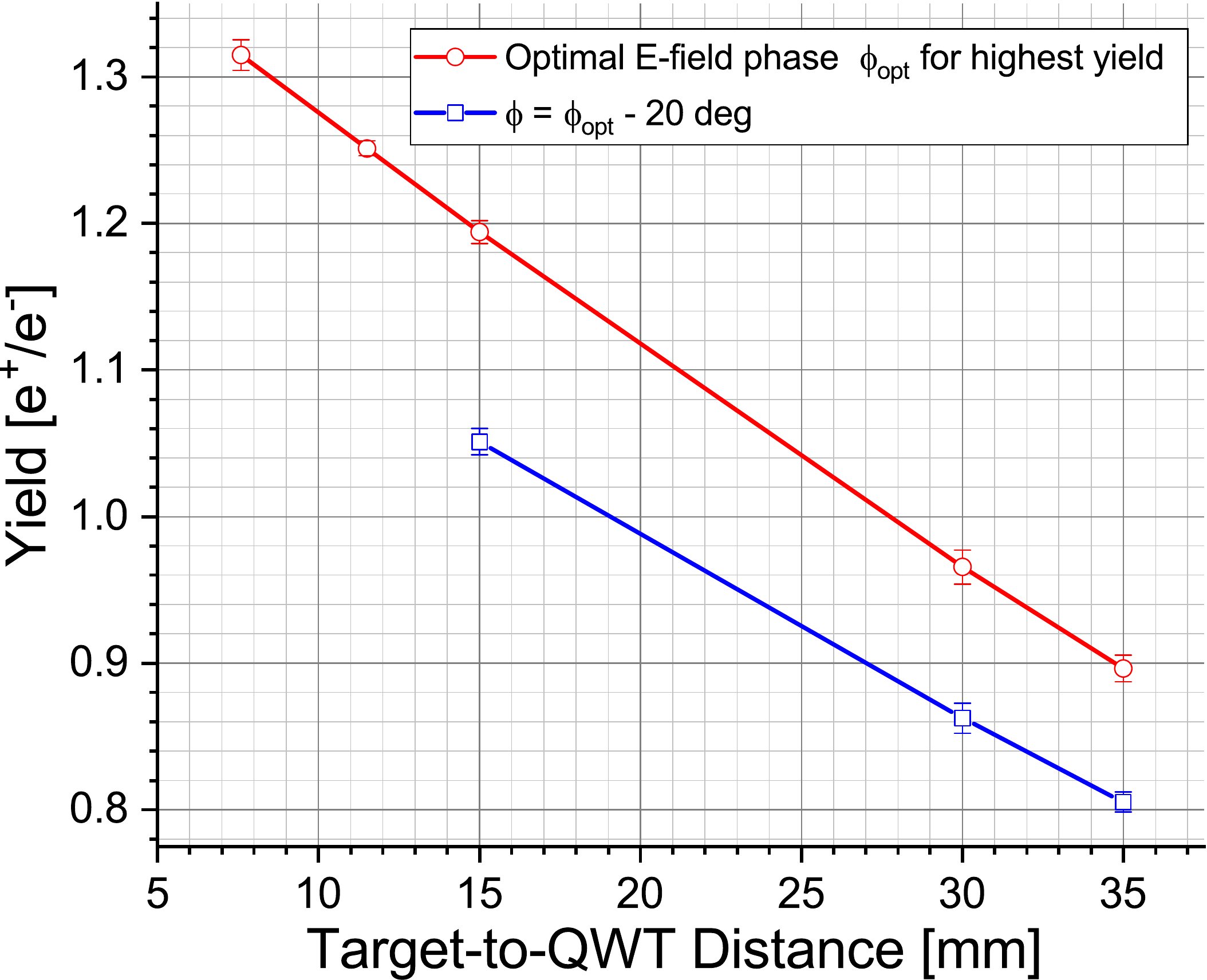}
\end{tabular}
  \caption{{\it{Left:}} Shape of the magnetic field on axis of the QWT. The yellow line shows the field as used for the simulation studies presented in this paper to achieve the required positron yield. The black line gives the B field as suggested for the QWT in reference~\cite{ref:GaiLiu-OMD}. The margenta line is the corresponding approximation used for the simulation of the poitron yield. 
{\it{Right:}} Positron yield depending on the distance d$_\mathrm{t2qwt}$ for optimized and non-optimized phase. The undulator $K$ value is $K=0.85$.}
 \label{fig:QWT-Bz-Y}
\end{figure}

During the long time of desgn studies for the ILC positron source several options for the OMD have been considered. An overview is given in reference~\cite{ref:GaiLiu-OMD}.  
The best solution would be a pulsed flux concentrator which initially  was envisaged for the adiabatic matching device (AMD). 
But in further detailed studies it appeared that over the long pulses of about 1\,ms the field in the FC could not be kept stable in time~\cite{ref:e+WG}.
 Further, studies~\cite{ref:AU-OMD} showed that the peak energy deposition in the center at the front of the FC is too high for ILC250. Neither a larger FC aperture nor a shorter distance of FC to  undulator could improve the situation substantially. 
 Therefore, further studies were pursued with a QWT. 
Following reference~\cite{ref:GaiLiu-OMD}, only a yield  $Y\le 1\,$e$^+/$e$^-$ can be reached for electron beam energies of 125\,GeV. Studies~\cite{ref:AU-OMD} showed that for the thinner conversion target, a maximum K value ($K=0.92$) and an optimized B field the yield can be increased to the required value. Figure~\ref{fig:QWT-Bz-Y} demonstrates the influence of the magnetic field shape on the positron yield. If the magnetic field rises from almost zero at the target exit within 8\,mm to a maximum value of 1.04\,T, a yield of about 1.5e$^+$/e$^-$ can be reached for high $K$ values. 
It is the question whether a QWT with the corresponding optimum magnetic field is possible, \ie{} maximum field of $\approx 1\,$T at a distance of 7--8\,mm from the target and almost zero at the target. A magnetic field on the target creates eddy currents which heat the target. Even if the heating is not important, brake effects have to be conserdered and taken into account for the final design and construction of the wheel.  

\subsection{Magnetic field in target}\label{sec:eddy}
Studies have shown that a high magnetic field at the target increases substantially the positron yield. However, the drawback of a high field are the eddy currents created in  the spinning target.
 They induce additional heat deposition and stress  and brake the target rotation. In addition, also the magnetic field in the target depends on the eddy current and  influences the positron distribution at the target exit. This effect has not been considered in the simulations.

The impact  of eddy currents on a target wheel 
 has been studied with simulations and in experiments~\cite{ref:UK-eddy,ref:ANL-eddy}. 
The experiment described in reference~\cite{ref:UK-eddy} used a wheel prototype made  of 1.56\,mm thick Ti6Al4V rim with spokes. The torque associated with eddy current production was measured for a range of  immersion  depths  and  magnetic flux  densities.    
The  measured  torque values correspond to  heat loads of up to 4.7\,kW when operating in fields of approximately 1\,T at 1500\,rpm, extrapolating to 8.0\,kW at 2000\,rpm. Based on this result it was recommended to keep the magnetic field at 0.5\,T or below at the  ILC target exit.
\\
If the magnetic field is pulsed the loads decrease enormously:  Since the pulses are short and the repetion rate is 5\,Hz,  the average power is reduced to only 1-2\% compared to a DC magnet. 
Only very recently a proposal has been made~\cite{ref:sievers-posipol18} 
to use a pulsed solenoid, similar to that used for LEP and other positron
sources, which with sufficiently long pulses of at most 4\,ms duration can 
provide a field of up to 4\,T and stable in time over the duration of the beam 
pulse of 1\,ms. Increase of positron yield could be within reach. 
\\
The pulsed operation would lead also to an acceptable heat load from the solenoid into the rotating 
wheel by induced currents inside the target. These currents could be further 
cut by laminating the target rim, by introducing radial cuts into the rim of the 
wheel. These studies will be pursued in the next future.

\section{Optimized undulator parameters for ILC250}\label{sec:undulator-par}

In case that the  QWT is the  OMD, the positron yield could be too low, \ie{} $Y\le 1\,$e$^+/$e$^-$. A longer undulator is not recommended. One reason is the photon beam emission angle: The opening angle of the photon beam  is proportional to $1/\gamma$, \ie{} for ILC250 a relatively high energy deposition could be possible  in the walls of a long superconducting undulator. To protect the undulator walls and the vacuum, masks will be inserted, preferably at the position of the quadrupoles~\cite{ref:adriana}. However, for long undulators, the modules at the end suffer from  photons with energies of few MeV (see more details in reference~\cite{ref:khaled-lcws18}. To absorb these photons, relatively long masks are required which also have to prevent that shower particles from the masks  enter the undulator aperture and degrade the vacuum. This problem is  under study.  \\
It could be more efficient to optimize the undulator and  photon beam  parameters for the ILC250 option.
The reduction of the undulator period to 10.5\,mm or even 10\,mm increases the energy of the photons (see equation~\ref{eq:Egamma}). 
The cross section of the pair production process increases with the photon energy; for example, in Ti the pair production cross section at 10\,MeV is about 25\% higher than at 7.5\,MeV. 
Lower periods imply lower $K$ values for constant $B_0$, and smaller $K$ values reduce the number of photons, see equation~(\ref{eq:Ngamma}). However, the higher photon energy increases the rate of pair production in the target and could compensate the lower photon number so that the required positron yield can be reached even for a shorter undulator. 
\\
Rough estimates demonstrated that for ILC250 the active length of the undulator could be reduced to  about 200\,m for $\lambda_\mathrm{u} = 10.5\,$mm and to 180\,m for  $\lambda_\mathrm{u} = 10.0\,$mm. 
\\
In general, with lower $K$ values the relative  contribution of higher harmonics goes down, and  the photon distribution is more focused to the axis. This also could reduce the power deposition in the undulator walls. To re-optimize the undulator parameters detailed studies will include the realistic undulator modules with errors in B field and period. In addition, 
the target thickness should be adopted accordingly to find the optimum choice for the positron yield, the load  and the  cooling efficiency for the target. 
\section{Summary}\label{sec:sum}

Studies are performed  to design the undulator-based ILC positron source which will 
generate a polarized positron beam. 
The design studies have made considerable progress over the past 
years, no show stoppers have appeared. Stategies for technical developments, 
prototyping and laboratory tests have been laid out (see, for instance, 
references~\cite{ref:e+WG,ref:sievers-posipol18,ref:SR-miniWS17}. Thus this 
design can be considered as solid and is to be pursued rigorously in the next future.

Main issue is the positron target wheel cooled by thermal radiation. 
It was shown that such a target will work. However, also the positron capture, 
in particular the OMD, are crucial for the positron source performance and need still R\&D work. 
All studies  done so far base on the helical undulator prototype~\cite{ref:undproto}. 
Further optimization of the undulator parameters could help to improve substantially 
the positron source. 
The almost final  paramemeters for all components of the positron soure, \ie{} undulator, 
target and OMD should be fixed as soon as possible to develop also the engineering design 
for a  full functioning target wheel and positron capture complex.  

\section*{Acknowledgment}
This work was supported by the German Federal Ministry of Education and Research,
Joint Research Project R\&D Accelerator ``Positron Sources'', Contract Number 05H15GURBA.

\end{document}